\documentclass[]{tPHM2e}

\newcommand{\nph}[1]{#1$p$#1$h$}

\begin{document}

\title{Coupled-cluster theory of a gas of strongly-interacting fermions in the dilute limit}

\author{Bogdan Mihaila$^{\rm a}$$^{\ast}$
        \thanks{$^\ast$Corresponding author. Email: bmihaila@lanl.gov}
        and Andres Cardenas$^{\rm a,b}$
        \\
        $^{\rm a}$\em{Los Alamos National Laboratory, Los Alamos, NM 87545};
        \\ 
        $^{\rm b}$\em{Mathematics Department, Cal Poly Pomona, Pomona, CA 91768}
       }

\maketitle

\begin{abstract}
We study the ground-state properties of a dilute gas of strongly-interacting fermions in the framework of the coupled-cluster expansion (CCE). We demonstrate that properties such as universality, opening of a gap in the excitation spectrum and applicability of s-wave approximations appear naturally in the CCE approach. In the zero-density limit, we show that the ground-state energy density depends on only one parameter which in turn may depend at most on the spatial dimensionality of the system.\bigskip

\begin{keywords}dilute fermion systems; electron gas; equation of state; unitarity limit.
\end{keywords}\bigskip

\end{abstract}


%
%

\section{Introduction}

One of the major endeavors in modern physics is the quest to isolate the underlying physics in a given system against the background of irrelevant complexity. Arguably, this challenge is tied into the ability to investigate systematically the predictions of a given theoretical Hamiltonian model. Separating model features from artifacts of the theoretical approximations involved, represents the key to the quantitative understanding of the physics of strongly-interacting systems. Unfortunately, first-principles calculations are presently possible only for small systems.

To achieve predictive power in many-body theory, it is important to develop a theoretical framework that can provide a systematic way of improving over mean-field theory and to demonstrate conclusively the numerical convergence of the results with the order of the approximation. For strongly-interacting systems of particles, traditional perturbation theory fails to converge fast enough for practical purposes. Hence, the desired nonperturbative character of the theoretical framework to be employed. In this context, we submit that the coupled-cluster expansion approach to solving the many-body Schr\"odinger equation represents the best hope of extending present state-of-the-art first-principles calculations to the realm of systems with large number of particles.

It is also important to identify problems involving strongly-interacting systems of particles that can provide the test bed for many-body theoretical approaches and approximations. One such system is the infinite (matter-like) many-body system composed of spin-1/2 fermions interacting via a zero-range, infinite scattering length contact interaction. The problem of finding the ground-state properties of this system is referred to sometimes as the ``George Bertsch problem''~\cite{BP,baker,bertsch,bulgac} and is of particular interest in astrophysics in connection with the equation of state for neutron matter. This problem has been revisited recently with the advent of experimental studies in ultracold fermionic atom gases of the crossover from the regime of Bardeen-Schriffer-Cooper (BCS) weakly-bound Cooper pairs to the regime of Bose-Einstein condensation (BEC) of diatomic molecules~\cite{ref:exp_Duke1,ref:exp_Duke2,ref:exp_Duke3,ref:exp_ENS,ref:exp_JILA1,ref:exp_JILA2,ref:exp_JILA3,ref:exp_MIT1,ref:exp_MIT2,ref:exp_MIT3,ref:exp_Innsbruck}.

The ``Bertsch problem'' was originally intended as a \emph{challenge} parameter-free model of neutron matter at subnuclear density. In cold atom physics, the regime of interest is known as the ``unitarity limit''~\cite{unitarity,unitarity_new}, i.e. the limit near Feshbach resonances where the s-wave scattering length, $a_0$, of two atoms with different spin components is much larger than the inter-particle distance ($k_\mathrm{F} |a_0| >> 1$). Here, $k_\mathrm{F}$ denotes the Fermi momentum of the gas, which is conventionally related to the total density of particles, $\rho_0$, by the noninteracting Fermi gas formula
\begin{equation}
   \rho_0 =
   \sum_{\sigma}
   \int_{\le k_\mathrm{F}} \frac{\mathrm{d}^3k}{(2\pi)^3}
   =
   k_\mathrm{F}^3 / (3 \pi^2)
   \>,
\end{equation}
where the momentum integral is performed over the interior volume of the Fermi sphere, and $\sigma$ denotes the spin component of the fermion, i.e. $\sigma = \pm \frac{1}{2}$.

The ``unitarity limit" corresponds to the BCS to BEC crossover in dilute ultracold fermionic atom gases and can be reached by modifying either the s-wave scattering length or the system density. For an interaction-induced crossover, the unitarity limit corresponds to the singularity in the scattering length and the limit is the same when approached with positive or negative scattering length. In this limit, the correlations are deemed to be significant and the system is expected to exhibit universal behavior, independent of the shape of the potential and dependent only on the particle density.

The importance of correlations in the ground state of dilute fermionic matter in the unitarity limit is measured by the numerical value of the ratio
\begin{equation}
   \xi = \varepsilon / \varepsilon_0
   \>,
\end{equation}
where $\varepsilon$ and $\varepsilon_0$ denote the ground-state energy densities of the interacting and noninteracting systems, respectively. A close upper-bound to the value of $\xi$ was set by the quantum Monte Carlo (QMC) study performed by Carlson \emph{et al}.~\cite{carlson}, which gave the value $\xi_\mathrm{QMC}=0.44 \pm 0.01$. In contrast, the ``universal'' curve describing the BCS to BEC crossover in the standard BCS variational picture~\cite{ref:BCS1,ref:BCS2} derived by Leggett~\cite{ref:leg80} gives the mean-field numerical value $\xi_{MF}=0.59$, which suggests that beyond-to-leading order effects are responsible for a change of about 25\% when compared to the mean-field solution.
Recent theoretical and experimental values for $\xi$ are summarized in Ref.~\cite{ref:th}.

In this paper, we will argue that by applying the coupled-cluster expansion (CCE)~\cite{ref:CCE1,ref:CCE2} formalism to the Bertsch problem, we can demonstrate conclusively that the resulting CCE equations are consistent with expectations based on purely heuristic arguments, such as universality and the possibility of a gap in the single-particle (s.p.) excitation spectrum. The latter is important for the ability to capture the superfluid properties of the system, if present. We will show that in the combined unitarity and zero-density limit, the ratio $\xi$ depends on only one parameter which in turn may depend only on the spatial dimensionality of the system.

%
%

\section{Coupled-cluster formalism}

We will review now the basics of the CCE approach: Consider the many-body Schr\"odinger equation, $\mathbf{H} \ | \Psi \rangle = E \ | \Psi \rangle$, with a two-body Hamiltonian \emph{density} operator, $\mathbf{H} = \mathbf{T} + \mathbf{V}$.

In second quantization, operators are expressed in terms of the creation and annihilation operators in momentum representation, ${\mathbf c}^\dag(\mathbf{q})$ and ${\mathbf c}(\mathbf{q})$, subject to the canonical relations:
\begin{align}
    \bigl \{ {\mathbf c}^\dag(\mathbf{q}_1), {\mathbf c}(\mathbf{q}_2) \bigr \}
    = \delta(\mathbf{q}_1 - \mathbf{q}_2)
    \>,
    \quad
    \bigl \{ {\mathbf c}^\dag(\mathbf{q}_1), {\mathbf c}^\dag(\mathbf{q}_2) \bigr \}
    =
    \bigl \{ {\mathbf c}(\mathbf{q}_1), {\mathbf c}(\mathbf{q}_2) \bigr \}
    = 0
    \>,
\end{align}
where spin (and isospin) degrees of freedom are implied. The
coordinate and momentum space representations of the creation
operators are related via the symmetric Fourier transform,
\begin{equation}
   \psi(\mathbf{x})
   = \frac{1}{(2\pi)^{3/2}}
   \int {\rm d}^3q \
        {\mathbf c}(\mathbf{q}) \, e^{{\rm i} \mathbf{q} \cdot \mathbf{x}}
   \>.
\end{equation}
Next, we introduced the particle and hole operators
\begin{equation}
   {\mathbf a}(\mathbf{q}) =
   {\mathbf c}(\mathbf{q}) \, \theta(q - k_\mathrm{F})
   \>,
   \quad
   {\mathbf b}(\mathbf{q}) =
   {\mathbf c}^\dag(\mathbf{q}) \, \theta(k_\mathrm{F} - q)
   \>,
\end{equation}
such that the physical vacuum obeys the relations
\begin{equation}
   {\mathbf a}(\mathbf{q})\ | \Phi \rangle = 0
   \>,
   \qquad
   {\mathbf b}(\mathbf{q})\ | \Phi \rangle = 0
   \>.
\end{equation}
With these definitions, we have
\begin{equation}
    {\mathbf c}^\dag(\mathbf{q}) =
    {\mathbf a}^\dag(\mathbf{q}) + {\mathbf b}(\mathbf{q})
    \>,
    \qquad
    {\mathbf c}(\mathbf{q}) =
    {\mathbf a}(\mathbf{q}) + {\mathbf b}^\dag(\mathbf{q})
    \>.
\end{equation}

In order to derive the CCE equations, we begin with the following ansatz for the many-body wave function, $| \Psi \rangle = e^\mathbf{S} \ | \Phi \rangle$, where $| \Phi \rangle$ is the \emph{physical} vacuum, and $\mathbf{S}$ is the many-body cluster correlation operator defined as
\begin{equation}
   \mathbf{S} = \mathbf{S}_1 + \mathbf{S}_2 + \mathbf{S}_3 + \cdots
   \label{s_eq}
   \>,
\end{equation}
where $\mathbf{S}_n$ gives rise to the \nph{n}-configuration contributions in $| \Psi \rangle$, i.e.
\begin{align}
   {\mathbf S_n}
   = &
   \frac{1}{n!}
   \int_{\ge k_\mathrm{F}} \!\! \mathrm{d}^3 p_1 \cdots
   \int_{\ge k_\mathrm{F}} \!\! \mathrm{d}^3 p_n
   \int_{\le k_\mathrm{F}} \!\! \mathrm{d}^3 k_n \cdots
   \int_{\le k_\mathrm{F}} \!\! \mathrm{d}^3 k_1
   \notag \\ & \ \times
        S_n(\mathbf{p}_1, \cdots \mathbf{p}_n; \mathbf{k}_n, \cdots \mathbf{k}_1) \,
        {\mathbf a}^\dag_{\mathbf{p}_n} \cdots {\mathbf a}^\dag_{\mathbf{p}_1}
        {\mathbf b}^\dag_{\mathbf{k}_1} \cdots {\mathbf b}^\dag_{\mathbf{k}_n}
   \label{sn_eq}
   \>.
\end{align}
By construction, the many-body wave function, $| \Psi \rangle$, obeys the normalization condition, $\langle \Psi | \Phi \rangle = 1$.

In the case of an infinite system, the physical vacuum, $| \Phi \rangle$, is represented by the noninteracting Fermi gas system. Because of translational invariance arguments, the s.p. representation of the vacuum is introduced in terms of plane-wave wave functions and the amplitudes $S_n(\mathbf{p}_1, \cdots \mathbf{p}_n; \mathbf{k}_n, \cdots \mathbf{k}_1)$ in $\mathbf{S}_n$, see Eq.~\eqref{sn_eq}, satisfy the property
\begin{equation}
   S_n(\mathbf{p}_1, \cdots \mathbf{p}_n; \mathbf{k}_n, \cdots \mathbf{k}_1)
   \propto
\label{eq:sn_inv}
   \delta^3 \Bigl ( \sum_{i=1}^n \mathbf{p}_i - \sum_{i=1}^n \mathbf{k}_i \Bigr )
   \>.
\end{equation}
It follows immediately that the \nph{1}-correlations term in~$\mathbf{S}$ vanishes, i.e. $\mathbf{S}_1 \equiv 0$.

Using the CCE ansatz, the Schr\"odinger equation can be written in normal ordered form, as
\begin{equation}
   \frac{1}{\rho_0} \,
   \bigl [ e^{- \mathbf{S}} \mathbf{H} e^\mathbf{S} \bigr ]_{c} \ | \Phi \rangle
   = \varepsilon \, | \Phi \rangle \>,
   \label{SE_of}
\end{equation}
where the subscript $c$ indicates the creation part of a normal-ordered operator. For $c=0$, Eq.~\eqref{SE_of} gives the expression for the ground-state energy density, which, in the case of a two-body Hamiltonian, i.e. if $\mathbf{V} \equiv \mathbf{V}^{(2)}$, includes at most contributions due to \nph{2}-correlations. For $c \neq 0$ Eq.~\eqref{SE_of} gives rise to a system of nonlinear equations that must be solved self-consistently for the amplitudes $S_n(\mathbf{p}_1, \cdots \mathbf{p}_n; \mathbf{k}_n, \cdots \mathbf{k}_1)$ in $\mathbf{S}_n$. In the case of an infinite system, the equation corresponding to $c=1$ is identically zero because $\mathbf{S}_1\equiv 0$. Furthermore, approximations to the CCE system of nonlinear equations are based on the idea that the $S_n$ amplitudes are small as long as the relative distance between particles is much larger than their average distance. Hence, approximations such as $\mathbf{S}_n \equiv 0$ for $n \ge N$, are equivalent to truncating an expansion in powers of density~\cite{FC:lecture}. Therefore, in the dilute limit, we can disregard contributions due to $\mathbf{S}_n$ with $n\ge 3$, and Eq.~\eqref{s_eq} gives simply $\mathbf{S} = \mathbf{S}_2$. We conclude that, for an infinite matter system of fermions in the dilute limit, we only need to solve the equations for $c=0$ (the energy-density equation) and $c=2$ (the equation for the amplitudes in $\mathbf{S}_2$).

%
%

\section{Interaction model}

Without loss of generality, we consider here the same two-body potential used by Carlson \emph{et al}.~\cite{carlson} in their QMC study of the unitarity limit, i.e.
\begin{equation}
   V(r) =
   - \frac{1}{m} \ \frac{\alpha^2}{\mathrm{cosh}^2(\alpha r)}
   \label{eq:cosh}
   \>.
\end{equation}
By construction, the \emph{reduced} wave function at zero energy in this potential is
\begin{equation}
   u_0(r) =
   r \, \psi_0(r)
   =
   \mathrm{tanh}(\alpha r)
   \>.
\end{equation}
It follows that the potential~\eqref{eq:cosh} corresponds to an infinite s-wave scattering length, $a_0=-\, \infty$. We find the range of the potential~\eqref{eq:cosh} is
\begin{equation}
   r_0 =
   \int_0^\infty \mathrm{d}r \, \phi_0(r) \,
   \bigl [ 2 - \phi_0(r) \bigr ]
   \equiv \frac{2}{\alpha}
   \>,
\end{equation}
where we have introduced the notation, $\phi_0 = 1 - u_0(r)$, to denote the difference between $u_0(r)$ and its asymptotic behavior. Finally, we note that the dilute limit is achieved by taking the limit $\alpha r_s \rightarrow \infty$. Here, $r_s$ denotes the unit radius defined as
\begin{equation}
   \rho_0 \, \frac{4\pi}{3} \, r_s^3 = 1
   \>.
\end{equation}
We obtain $r_s = 1 / (\gamma k_\mathrm{F})$, with $\gamma = \bigl[ 4/(9\pi) \bigr ]^{1/3}   $.

We will write the energy density of fermionic matter in terms of the ratio
\begin{equation}
   \frac{V(y/k_\mathrm{F})}{\varepsilon_0}
   =
   - \frac{10\, \gamma^2}{3} \,
     \frac{(\alpha r_s)^2}{\mathrm{cosh}^2\bigl [ y \,
        \gamma (\alpha r_s) \bigr ]}
   \label{eq:vpot_y}
   \>,
\end{equation}
where we have introduced the notation $y=k_\mathrm{F} \, r$. With our choice of potential, see Eq.~\eqref{eq:cosh}, the ratio $\xi$ depends on the product $\alpha r_s$, where $\alpha$ is a measure of the range of the interaction and $r_s$ is related to the Fermi gas density, as described above. We note that Carlson \emph{et al}. have carried out QMC calculations for $\alpha r_s$ values of 12 and 24, and have noted that the changes in $\xi$ between these two values of $\alpha r_s$ were negligible within statistical errors.

%
%

\section{Energy density in the Bertsch problem}

\begin{figure}[t!]
   \centering
   \includegraphics[width=3in]{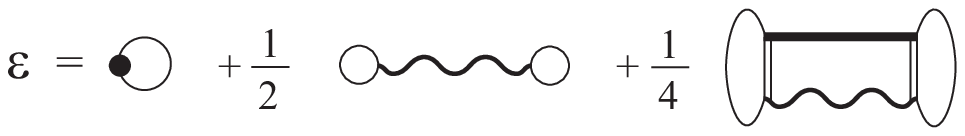}
   \caption{Diagrammatic representation of the energy of an infinite two-body Hamiltonian system.
   Here, vertical single and double lines depict hole and particle states, respectively,
   i.e. states with momenta $k\le k_\mathrm{F}$ and $p \ge k_\mathrm{F}$, respectively;
   filled circles indicate the kinetic energy operators,
   whereas horizontal wiggles and thick lines indicate the two-body potential operator and
   the \nph{2}-correlation amplitudes, $S_2(\mathbf{p}_1, \mathbf{p}_2; \mathbf{k}_2, \mathbf{k}_1)$,
   respectively.
   }
   \label{fig1:erg}
\end{figure}

We recall that, in the case of an infinite matter system, we have $\mathbf{S}_1=0$. Then, the energy density reads
\begin{equation}
   \varepsilon
   =
   \frac{1}{\rho_0} \,
   \langle \Phi |
   e^{-\mathbf{S}} {\mathbf H} e^\mathbf{S}
   | \Phi \rangle
\label{eq:erg_den}
   =
   \frac{1}{\rho_0} \,
   \langle \Phi |
   \Bigl \{
      \mathbf{T} +
      \mathbf{V} +
      \bigl [ \mathbf{V}, \mathbf{S}_2 \bigr ]
   \Bigr \}
   | \Phi \rangle
   \>.
\end{equation}
This equation is \emph{exact} and is illustrated diagrammatically in Fig.~\ref{fig1:erg}.

%
%

\subsection{Vacuum expectation of the kinetic energy}

The first term in Eq.~\eqref{eq:erg_den} represents the expectation value of the kinetic-energy operator in the vacuum state. As stated above, in the case of infinite matter, the vacuum is the noninteracting Fermi gas. Hence, we have
\begin{equation}
   \varepsilon_0
   = \,
   \frac{1}{\rho_0} \
   \langle \Phi | {\mathbf T} | \Phi \rangle
   = \,
   \sum_{\sigma}
   \int_{\le k_\mathrm{F}} \frac{\mathrm{d}^3k}{(2\pi)^3} \
   \langle \mathbf{k} | \Bigl ( - \frac{ \nabla^2 }{2 m} \Bigr ) | \mathbf{k} \rangle
   \>,
\end{equation}
which gives the familiar result, $ \varepsilon_0 = (3/5) \varepsilon_\mathrm{F}$, with $\varepsilon_\mathrm{F} = k_\mathrm{F}^2 / (2m)$. This allows one to write the \emph{exact} expression for the ratio $\xi$, as
\begin{equation}
   \xi
\label{eq:cce_erg}
   =
   1 +
   \frac{1}{\rho_0 \varepsilon_0} \,
   \langle \Phi |
   \Bigl \{
   \mathbf{V}
   + \bigl [ \mathbf{V}, \mathbf{S}_2 \bigr ]
   \Bigr \}
   | \Phi \rangle
   \>.
\end{equation}
Therefore, the quantity ($\xi$-1) measures the departure from the noninteracting Fermi gas result. This is given by the sum of two contributions: i) the expectation value of the two-body interaction operator, $\mathbf{V}$, in the vacuum state, and ii) the expectation value of the commutator $\bigl [ \mathbf{V}, \mathbf{S}_2 \bigr ]$ in the vacuum state. (We note that, by construction, we have
$
   \langle \Phi | \bigl [ \mathbf{V}, \mathbf{S}_2 \bigr ] | \Phi \rangle
   =
   \langle \Phi | \mathbf{V} \, \mathbf{S}_2 | \Phi \rangle
$.)

%
%

%
%

\begin{figure*}[t]
   \centering
   \includegraphics[width=\textwidth]{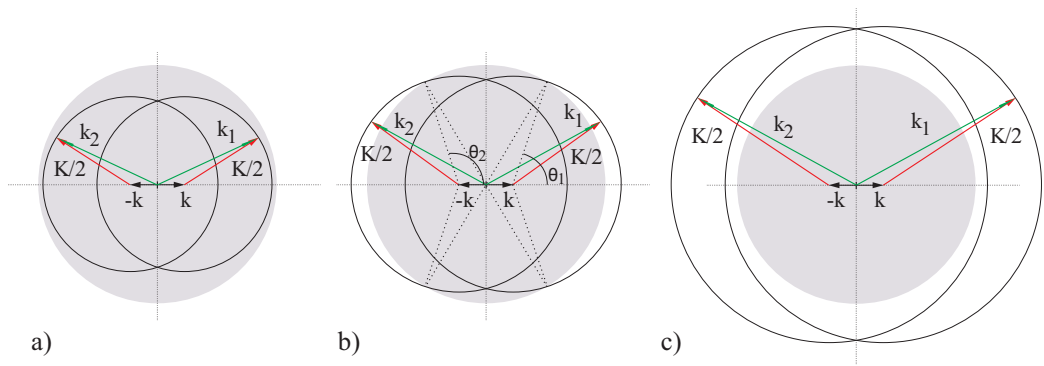}
   \caption{(Color online) The integration domain of the center-of-mass coordinate, $\mathbf{K}$, depends on the value of the magnitude of the relative coordinate, $k$, as follows:
   a) $\frac{1}{2}K \le k_\mathrm{F} - k$,
   b) $k_\mathrm{F} - k \le \frac{1}{2}K \le \sqrt{k_\mathrm{F}^2 + k^2}$,
   c) $\sqrt{k_\mathrm{F}^2 + k^2} \le \frac{1}{2}K$.
   The shaded areas indicate the Fermi spheres ($k\le k_\mathrm{F}$).
   }
   \label{fig2:DK}
\end{figure*}

%
%

\subsection{Vacuum expectation of the potential }

We discuss now the contribution of the two-body potential, $\mathbf{V}$, alone. We have
\begin{equation}
   \langle \Phi | \mathbf{V}
   | \Phi \rangle
   \label{eq:vpot_d}
   =
   \frac{1}{2}
   \sum_{S=0,1} (2S+1)
   \int_{\le k_\mathrm{F}} \frac{\mathrm{d}^3k}{(2\pi)^3}
   \int_{\mathcal{D}_K} \frac{\mathrm{d}^3 K}{(2\pi)^3} \
      \langle \mathbf{k} | \mathbf{V}^{(S,a)} | \mathbf{k} \rangle
   \>,
\end{equation}
where $S$ denotes the total spin of a pair of fermions, $\mathbf{k}$ and $\mathbf{K}$ represent the relative and center-of-mass coordinates of the pair, respectively, and the \emph{anti-symmetric} matrix element of the interaction is defined as
\begin{align}
   \langle \mathbf{q}' | \mathbf{V}^{(S,a)} | \mathbf{q} \rangle
   =
   \langle \mathbf{q}' | \mathbf{V}^{(S)} | \mathbf{q} \rangle
   + (-1)^S
   \langle -\mathbf{q}' | \mathbf{V}^{(S)} | \mathbf{q} \rangle
   \>.
\end{align}
For a central potential, $\mathbf{V} \equiv V(r)$, the matrix element of the interaction reads
\begin{equation}
   \langle \mathbf{q}' | \mathbf{V}^{(S)} | \mathbf{q} \rangle
   =
   \int \mathrm{d}^3r \ V(r) \ e^{\mathrm{i} (\mathbf{q}-\mathbf{q}') \cdot \mathbf{r}}
   \>.
\end{equation}

In Eq.~\eqref{eq:vpot_d}, $\mathcal{D}_K$ indicates the integration domain for the variable $\mathbf{K}$, which is illustrated in Fig.~\ref{fig2:DK}. Using
\begin{align}
   \int_{\mathcal{D}_K} \frac{\mathrm{d}^3 K}{(2\pi)^3}
   = &
   4 \, \rho_0 \,
   {\cal B}(k)
   \>,
\end{align}
where we have introduced the notation,
\begin{equation}
   {\cal B}(k)
   =
   1
   -
   \frac{3}{2} \, \Bigl ( \frac{k}{k_\mathrm{F}} \Bigr )
   +
   \frac{1}{2} \, \Bigl ( \frac{k}{k_\mathrm{F}} \Bigr )^3
   \>,
\end{equation}
we obtain
\begin{equation}
\label{eq:vpot}
   \frac{1}{\rho_0 \varepsilon_0} \,
   \langle \Phi | \mathbf{V}
   | \Phi \rangle
   =
   4
   \int_0^{k_\mathrm{F}} \mathrm{d}k \ k^2 \, {\cal B}(k)
   \int_0^\infty \mathrm{d}r \, r^2 \
   [ V(r)/\varepsilon_0 ] \,
   \bigl [ 2 - j_0(2 k r) \bigr ]
   \>.
\end{equation}
Here $j_\ell(z)$ denotes the regular spherical Bessel function of rank $\ell$.

%
%

\begin{figure}[b!]
   \centering
   \includegraphics[width=3.5in]{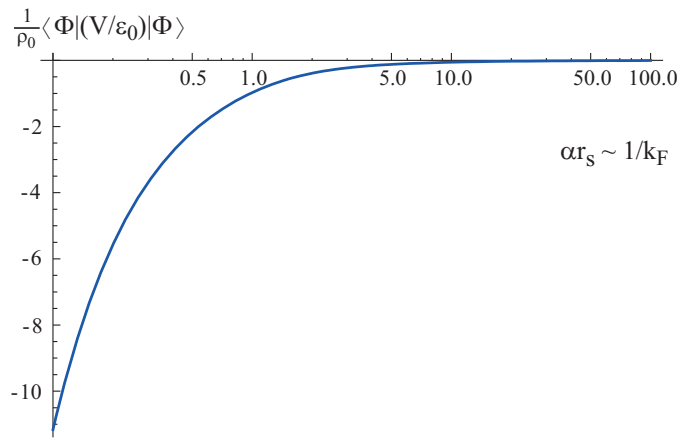}
   \caption{Vacuum expectation value of the two-body potential, with the
    inter-particle density given by the infinite s-wave scattering length potential
    introduced by Carlson \emph{et al}.~\cite{carlson}.
    Here, the dilute limit is recovered for $\alpha r_s \gg 1$.
   }
   \label{fig3:vpot}
\end{figure}

%
%

We note that, using the power expansion of $j_0(z)$, we can show that the radial integral in Eq.~\eqref{eq:vpot} can be calculated only in terms of the \emph{even} momenta of the potential, $\langle r^{2n} V(r) \rangle$, i.e.
\begin{equation}
   \int_0^\infty \mathrm{d}r \, r^2 \
   V(r) \,
   \bigl [ 2 - j_0(2 k r) \bigr ]
   \equiv
   \langle V(r) \rangle
   +
   \frac{2 k^2}{3} \ \langle r^2 V(r) \rangle
   -
   \cdots
\label{eq:v_mom}
   \>,
\end{equation}
with
\begin{align}
   \langle r^{2n} V(r) \rangle
   =
   \int_0^\infty [ \mathrm{d}r \, r^2 ] \ r^{2n} \, V(r)
   \>.
\end{align}

We evaluate the ratio $V(r)/\varepsilon_0$ using Eq.~\eqref{eq:vpot_y}. Then, the density dependence of the expectation value of the interaction in the vacuum, see Eq.~\eqref{eq:vpot}, is illustrated in Fig.~\ref{fig3:vpot}.  We notice that in the zero-density limit, this contribution to the energy density vanishes. Using the explicit form of the interaction, Eq.~\eqref{eq:vpot_y}, we can show that the momenta of the potential, $\langle r^{2n} V(r) \rangle$, decrease in size with increasing $n$, at fixed density. So, according to Eq.~\eqref{eq:v_mom}, in the dilute limit the contribution due to $\langle \Phi | \mathbf{V} | \Phi \rangle$ depends only on $\langle V(r) \rangle$, which is independent of the shape of the potential. Therefore, independent of the explicit choice of the inter-particle interaction, we obtain that the zero-density limit of the ratio $\xi$ in the CCE approach is obtained as
\begin{equation}
   \xi \rightarrow
   \xi_0 =
   1 +
   \frac{1}{\rho_0 \varepsilon_0} \,
   \langle \Phi |
   \bigl [ \mathbf{V}, \mathbf{S}_2 \bigr ]
   | \Phi \rangle
\label{eq:xi-0}
   \>.
\end{equation}
Hence, the \emph{universal} character of this result in the unitarity limit. Moreover, the quantity ($\xi_0$-1) represents a direct measure of the importance of correlations in the ground state.

%
%

\subsection{Correlations effects}

\begin{figure*}[t!]
   \centering
   \includegraphics[width=0.8\textwidth]{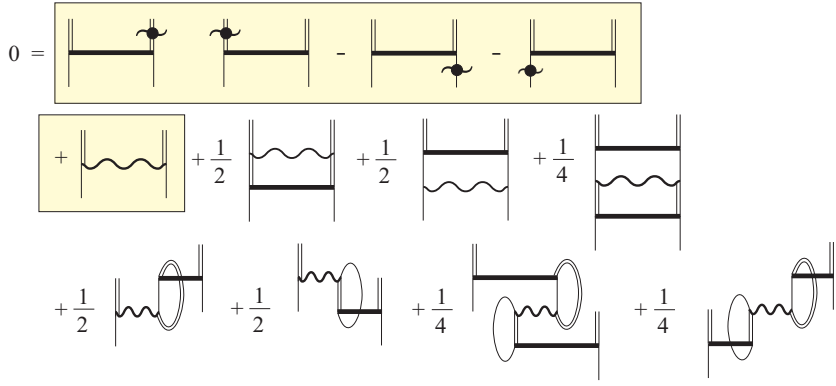}
   \caption{(Color online)
   Diagrammatic representation of the $\mathbf{S}_2$ equation in the case of a
   dilute infinite matter system of spin-1/2 fermions.
   The four terms boxed on the first line read as $S_2(\mathbf{p}_1, \mathbf{p}_2; \mathbf{k}_2, \mathbf{k}_1) \, (\tilde \epsilon_{p_1} + \tilde \epsilon_{p_2} - \tilde \epsilon_{k_1} - \tilde \epsilon_{k_2})$, where $\tilde \epsilon_{p \, (k)}$ are the renormalized s.p. energies given in Fig.~\ref{fig5:gap}.
   }
   \label{fig4:S2_eqn}
\end{figure*}

\begin{figure}[b!]
   \centering
   \includegraphics[width=3in]{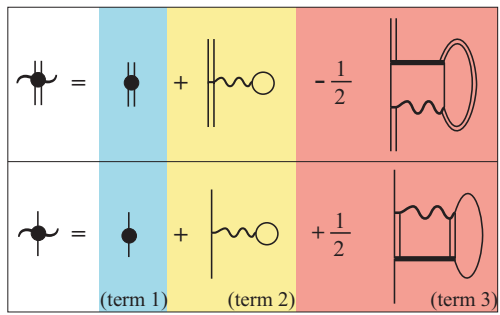}
   \caption{(Color online)
   Diagrammatic representation of the renormalized s.p. spectrum
   in the case of a dilute infinite matter system of spin-1/2 fermions.
   The first line gives the \emph{particle}-state s.p. energies, $\tilde \epsilon_p$,
   whereas the second line gives the \emph{hole}-state s.p. energies, $\tilde \epsilon_k$.
   In either case, the renormalized s.p. energies include the noninteracting s.p. energies,
   $\epsilon_p = p^2/(2m)$ and $\epsilon_k = k^2/2m$, corrected for interaction (terms 2 and 3)
   and correlations (term 3) effects. We note that terms 1 and 2 have the same signs for
   both particle and hole states, whereas terms 3 have opposite signs and give rise to
   a gap in the s.p. energy spectrum.
   }
   \label{fig5:gap}
\end{figure}

We turn now to the calculation of the correlations effects. We begin by recalling that according to Eq.~\eqref{eq:sn_inv} we can write
\begin{equation}
   S_2(\mathbf{p}_1, \mathbf{p}_2; \mathbf{k}_2, \mathbf{k}_1)
   \equiv
   Z( \mathbf{p}, \mathbf{k}; \mathbf{K} ) \
   \delta^3(\mathbf{P} - \mathbf{K})
\label{eq:s2_inv}
   \>.
\end{equation}
Here, $\mathbf{k}$ is confined to the volume of the Fermi sphere, i.e. $k \le k_\mathrm{F}$, $\mathbf{p}$ spans the entire space and the center-of-mass momentum $\mathbf{K}$ is confined to the domain illustrated in Fig.~\ref{fig2:DK}.
Then, we obtain
\begin{align}
   \label{eq:vs2_d}
   \langle \Phi | &
   \bigl [ \mathbf{V}, \mathbf{S}_2 \bigr ]
   | \Phi \rangle
   \\ \notag &
   =
   \frac{1}{2}
   \sum_{S=0,1} (2S+1)
   \int_{\le k_\mathrm{F}} \frac{\mathrm{d}^3 k}{(2\pi)^3}
   \int_{\mathcal{D}_K} \frac{\mathrm{d}^3 K}{(2\pi)^3}
   \int \!\! \mathrm{d}^3p \
      Z^{(S,a)}(\mathbf{p},\mathbf{k};\mathbf{K}) \
      \langle \mathbf{k} | \mathbf{V}^{(S)} | \mathbf{p} \rangle
   \>.
\end{align}
Next, we perform the partial wave decomposition of the amplitude $Z^{(S)}(\mathbf{p},\mathbf{k};\mathbf{K})$, i.e.
\begin{equation}
   Z^{(S)}( \mathbf{p}, \mathbf{k}; \mathbf{K} )
   =
   \sum_{\ell_1 \ell_2 L} Z^{(S)}_{\ell_1 \ell_2 L}(p,k;K) \,
   \biggl ( \Bigl [ Y^{(\ell_1)}(\hat p) \otimes Y^{(\ell_2)}(\hat k) \Bigr ]^{(L)}
            \odot Y^{(L)}(\hat K) \biggr )
   \>,
\end{equation}
where we denote by $Y^{(\ell)}(\hat q)$ the spherical harmonic of rank $\ell$ that depends on the angular coordinates of the vector $\mathbf{q}$. To calculate the expectation value in Eq.~\eqref{eq:vs2_d}, we need to solve the $\mathbf{S}_2$ equation illustrated diagrammatically in Fig.~\ref{fig4:S2_eqn}. This equation is nonlinear, as expected because of the nonperturbative character of the CCE approach. For illustrative purposes, we note the terms boxed in on the first two lines in Fig.~\ref{fig4:S2_eqn} lead to the familiar approximation
\begin{equation}
   S_2^{(S,a)}(\mathbf{p}_1, \mathbf{p}_2; \mathbf{k}_2, \mathbf{k}_1)
   =
   - \,
   \frac{ \langle \mathbf{p}_1 \mathbf{p}_2 | \mathbf{V}^{(S,a)} | \mathbf{k}_2, \mathbf{k}_1 \rangle }
        { \tilde \epsilon_{p_1} + \tilde \epsilon_{p_2} - \tilde \epsilon_{k_1} - \tilde \epsilon_{k_2} }
   + \cdots
\label{eq:s2_app}
   \>,
\end{equation}
where $\tilde \epsilon_{p \, (k)}$ are the renormalized s.p. energies illustrated in Fig.~\ref{fig5:gap}. If instead one uses the noninteracting s.p. energies, then the $S_2(\mathbf{p}_1, \mathbf{p}_2; \mathbf{k}_2, \mathbf{k}_1)$ amplitudes are singular when all momenta are located on the Fermi sphere. Fortunately, as seen from Fig.~\ref{fig5:gap}, the renormalized s.p. energies include renormalizations for interaction (terms 2 and 3) and correlations (term 3) effects. We note that terms 1 and 2 have the same signs for both particle and hole states, whereas terms 3 have opposite signs and give rise to a gap in the s.p. energy spectrum.

\begin{figure}[t!]
   \centering
   \includegraphics[width=3.2in]{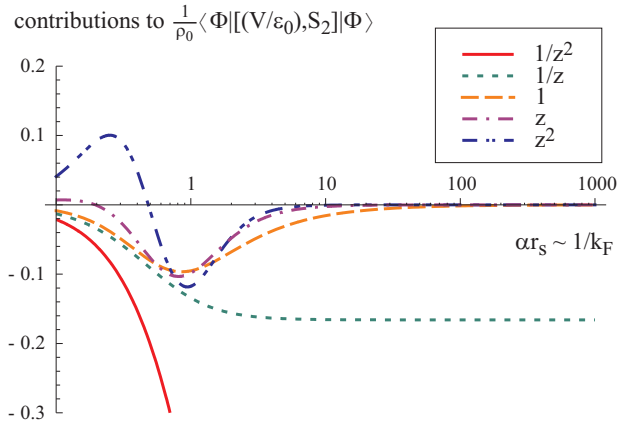}
   \caption{(Color online)
   Contributions to the vacuum expectation value, $\langle \Phi | \bigl [ \mathbf{V}, \mathbf{S}_2 \bigr ] | \Phi \rangle$, corresponding to powers $z^n$, $n=-2,\cdots 2$, in the power expansion of $\chi^{(0,a)}_{0 0 0}(z)$. Here, we have $z=kr$.}
   \label{fig6:vs2-zn}
\end{figure}

In order to reveal the true nature of the correlations effects in~\eqref{eq:xi-0}, it is necessary to introduce the 2-\emph{hole} function, defined as~\cite{c-CCE}
\begin{equation}
   \chi^{(S)}_{\mathbf{k} \mathbf{K}}(\mathbf{r})
   =
   \int {\rm d}^3 p \
   e^{{\rm i} \, \mathbf{p} \cdot \mathbf{r}} \
   Z^{(S)}(\mathbf{p}, \mathbf{k}; \mathbf{K})
   \>.
\end{equation}
This is equivalent to Fourier transforming the $\mathbf{p}$ degrees of freedom of the amplitudes $Z( \mathbf{p}, \mathbf{k}; \mathbf{K} )$. The associated partial-wave-expansion components of the 2-\emph{hole} function obey the relation
\begin{equation}
   \chi^{(S)}_{\ell_1 \ell_2 L}(r,k;K)
   =
   \int_0^\infty \!\! \mathrm{d}p \, p^2 \,
   j_{\ell_1}(pr) \,
   Z^{(S)}_{\ell_1 \ell_2 L}(p,k;K)
   \>,
\end{equation}
and have the anti-symmetric correspondent
\begin{equation}
   \chi^{(S,a)}_{\ell_1 \ell_2 L}(r,k;K)
   =
   \Bigl \{
      1 + \frac{1}{2} \bigl [ (-)^{S+\ell_1} + (-)^{S+\ell_2} \bigr ]
   \Bigr \} \
   \chi^{(S)}_{\ell_1 \ell_2 L}(r,k;K)
   \>.
\end{equation}
We obtain
\begin{align}
\label{eq:vs2}
   \frac{1}{\rho_0 \varepsilon_0} \,
   \langle \Phi |
   \bigl [ \mathbf{V}, \mathbf{S}_2 \bigr ]
   | \Phi \rangle
   = & \,
   \frac{2}{\pi^{3/2}} \
   \sum_\ell \ (-1)^\ell \, \sqrt{2\ell+1}
   \\ \notag & \times
   \int_0^{k_\mathrm{F}} \!\! \mathrm{d}k \, k^2 \,
   {\cal B}(k)
   \int_0^\infty \!\! \mathrm{d}r \, r^2 \,
   [ V(r)/\varepsilon_0 ] \,
   j_\ell(kr) \,
   \chi^{(S)}_{\ell \ell 0}(kr;0)
   \>,
\end{align}
where the spin component $S$ obeys the selection rule $\ell + S = \mathrm{even}$. From Eq.~\eqref{eq:s2_app}, $\chi^{(S,a)}_{\ell_1 \ell_2 L}(r,k;K)$ obeys an equation of the form
\begin{align}
   \frac{1}{m}
   \Bigl [ - \frac{1}{r^2}\frac{\mathrm{d}}{\mathrm{d}r}
     & \Bigl ( r^2 \frac{\mathrm{d}}{\mathrm{d}r} \Bigr )
     +
     \frac{\ell_1(\ell_1+1)}{r^2}
     -
     k^2
   \Bigr ] \,
   \chi^{(S,a)}_{\ell_1 \ell_2 L}(r,k;K)
   + \
   \cdots
   \\ \notag &
   =
   -
   (-1)^{\ell_1} \sqrt{4\pi (2\ell_1 + 1)} \
   \bigl [ 1 + (-1)^{S+\ell_1} \bigr ] \
   V(r) \ j_{\ell_1}(kr) \
   \delta_{\ell_1 \ell_2} \,
   \delta_{L0}
   \>,
\end{align}
which is similar to the radial part of a nonlinear scattering problem.

In particular, we are interested in the 2-\emph{hole} function partial-wave component, $\chi^{(S,a)}_{\ell \ell 0}(kr;0)$, that enters Eq.~\eqref{eq:vs2}. In order to write the general form of $\chi^{(S,a)}_{\ell \ell 0}(kr)$ it is useful to limit first the discussion to the case of the s-wave approximation of~\eqref{eq:vs2}, i.e. $\ell=S=0$, and consider the power expansion of $\chi^{(0,a)}_{0 0 0}(kr;0)$, i.e.
\begin{equation}
   \chi^{(0,a)}_{0 0 0}(z;0) = \sum_n \, c_n \, z^n
   \label{eq:powers}
   \>,
\end{equation}
where $z=kr$. In Fig.~\ref{fig6:vs2-zn} we depict the density dependence of the $z^n$ contributions to the vacuum expectation value, $\langle \Phi | \bigl [ \mathbf{V}, \mathbf{S}_2 \bigr ] | \Phi \rangle$. We notice that powers $z^n$ with $n \leq -2$ lead to divergent contributions, so only $n \geq -1$ are allowed in Eq.~\eqref{eq:powers}. Unlike the solution of a typical scattering problem, the solution $\chi^{(0,a)}_{0 0 0}(kr;0)$ is allowed an irregular part at the origin for fixed $k$. We also find that the powers corresponding to $n \geq 0$ vanish in the zero-density limit. Therefore the value of $(\xi_0-1)$ from Eq.~\eqref{eq:xi-0} is proportional only to the coefficient $c_{-1}$.

In general, the 2-\emph{hole} function partial-wave component, $\chi^{(0,a)}_{\ell \ell 0}(kr;0)$, that enters Eq.~\eqref{eq:vs2} has the form
\begin{equation}
   \chi^{(S,a)}_{\ell \ell 0}(kr;0)
   \label{eq:chi_exp}
   =
   A_\ell j_\ell(kr)
   +
   B_\ell n_\ell(kr) \, \delta_{\ell 0}
   +
   \bar \chi^{(S,a)}_{\ell \ell 0}(kr;0)
   \>.
\end{equation}
Solving the $\mathbf{S}_2$ equation is equivalent to finding the expansion coefficients $A_\ell$ and $B_\ell$ and the corrections $\bar \chi^{(S,a)}_{\ell \ell 0}(kr)$. Similar to our discussion surrounding Fig.~\ref{fig6:vs2-zn}, we can show by direct computation that the only allowed irregular contribution to $\langle \Phi | \bigl [ \mathbf{V}, \mathbf{S}_2 \bigr ] | \Phi \rangle$ corresponds to $\ell=0$. Hence, the irregular spherical component in Eq.~\eqref{eq:chi_exp}.

\begin{figure}[t!]
   \centering
   \includegraphics[width=3.2in]{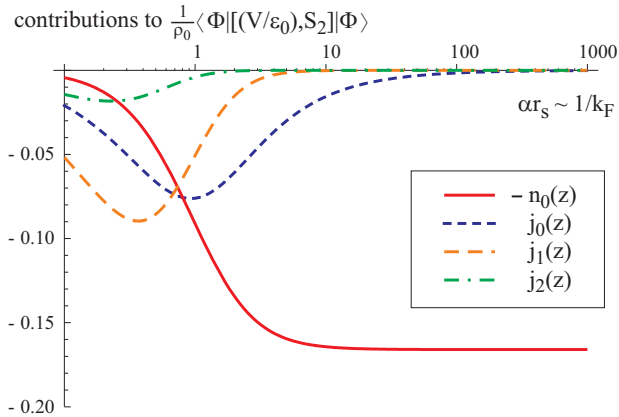}
   \caption{(Color online)
   Contributions to the vacuum expectation value, $\langle \Phi | \bigl [ \mathbf{V}, \mathbf{S}_2 \bigr ] | \Phi \rangle$, corresponding to the regular and irregular spherical Bessel components for ($\ell$=0,1,2).
   Here, we have $z=kr$.}
   \label{fig7:vs2-jl}
\end{figure}

In Fig.~\ref{fig7:vs2-jl} we illustrate the density dependence of the contributions to the vacuum expectation value, $\langle \Phi | \bigl [ \mathbf{V}, \mathbf{S}_2 \bigr ] | \Phi \rangle$, corresponding to the $n_0(k r)$ and the $j_\ell(k r)$ components, with $\ell$=0,1,2. We note that all $j_\ell(k r)$ components vanish in the zero-density limit, whereas the only nonzero contribution corresponds to the irregular s-wave spherical component, $n_0(k r)$. At low but finite density, e.g. $\alpha r_s > 5$, the contributions due the $j_\ell(k r)$ components decrease with increasing values of $\ell$. For $\alpha r_s > 10$, the dominant contributions correspond to the s-wave approximation, i.e.~$\ell=0$. Therefore, the value of $(\xi_0-1)$ from Eq.~\eqref{eq:xi-0} is proportional to the coefficient $B_0$. In this limit, the only scale present in the problem is the number of spatial of dimensions. Thus, the coefficient $B_0$ may depend at most on the spatial dimensionality of the system.

%
%

\section{Conclusions}

To summarize, in this paper we report results of a formal study of the ground-state properties of dilute fermionic matter in the unitarity limit, carried out using the CCE framework. In this approach, we are able to demonstrate properties of the strongly-interacting fermionic matter such as universality, the presence of a gap in the excitation spectrum and applicability of s-wave approximations in the dilute limit. We note that these results were obtained in an \emph{ab initio} fashion and did not invoke explicit gap parameters to be optimized variationally such as it is done in the BCS mean-field picture. This is particularly important, because it assures that a CCE-based description of the density-induced BCS to BEC crossover in the unitarity limit is indeed possible. In the zero-density limit, the ground-state CCE equations show that the ground-state energy density depends on only one parameter, which in turn may depend only on the spatial dimensionality of the system. In three spatial dimensions, the departure from the Fermi-gas energy-density result, measured by the numerical value of ($\xi_0$-1), is proportional to the expansion coefficient of $\mathrm{n}_0(kr)$, the only allowed irregular piece of the \emph{hole} function, $\chi^{(S,a)}_{\ell \ell 0}(kr)$. The \emph{hole}-function formalism introduced in Ref.~\cite{c-CCE} is the key ingredient that allows one to obtain this result.


\section*{Acknowledgements}

This work was supported by the Los Alamos National Laboratory under the auspices of the U.S. Department of Energy, under the LDRD program at Los Alamos National Laboratory. The authors gratefully acknowledge useful conversations with P.B.~Littlewood, J.L.~Smith, J.F.~Dawson and J.H.~Heisenberg.

%
%


\begin{thebibliography}{99}
   \bibitem{BP}
      R.F.~Bishop,
      Int. Mod. Phys. B \textbf{15}, Nos. 10 \& 11, iii (2001).

   \bibitem{baker}
      G.A.~Baker,
      Int. Mod. Phys. B \textbf{15}, Nos. 10 \& 11, 1314 (2001).

   \bibitem{bertsch}
      T.~Papenbrock and G.F.~Bertsch,
      Phys. Rev. C \textbf{59}, 2052 (1999).

   \bibitem{bulgac}
      A. Bulgac, J.E.~Drut, and P. Magierski,
      Phys. Rev. Lett. \textbf{96}, 090404 (2006).

   \bibitem{ref:exp_Duke1}
      K.M.~O'Hara, S.L.~Hemmer, M.E.~Gehm \textit{et al.},
      Science \textbf{298}, 2179 (2002).

   \bibitem{ref:exp_Duke2}
      M.E.~Gehm, S.L.~Hemmer, S.R.~Granade \textit{et al.},
      Phys. Rev. A \textbf{68}, 011401(R) (2003).

   \bibitem{ref:exp_Duke3}
      M.E.~Gehm, S.L.~Hemmer, K.M.~OHara \textit{et al.},
      Phys. Rev. A \textbf{68}, 011603 (2003).

   \bibitem{ref:exp_ENS}
      T.~Bourdel, J. Cubizolles, L. Khaykovich \textit{et al.},
      Phys. Rev. Lett. \textbf{91}, 020402 (2003).

   \bibitem{ref:exp_JILA1}
      C.A.~Regal, C.~Ticknor, J.L.~Bohn \textit{et al.},
      Nature (London) \textbf{424}, 47 (2003).

   \bibitem{ref:exp_JILA2}
      C.A.~Regal and D.S.~Jin,
      Phys. Rev. Lett. \textbf{90}, 230404 (2003).

   \bibitem{ref:exp_JILA3}
      C.A.~Regal, M.~Greiner, and D.S.~Jin,
      Phys. Rev. Lett. \textbf{92}, 040403 (2004).

   \bibitem{ref:exp_MIT1}
      S.~Gupta, Z.~Hadzibabic, M.W.~Zwierlein \textit{et al.},
      Science \textbf{300}, 1723 (2003).

   \bibitem{ref:exp_MIT2}
      S.~Gupta, Z.~Hadzibabic, J.R.~Anglin \textit{et al.},
      Phys. Rev. Lett. \textbf{92}, 100401 (2004).

   \bibitem{ref:exp_MIT3}
      M.W.~Zwierlein, C.A.~Stan, C.H.~Schunck \textit{et al.},
      Phys. Rev. Lett. \textbf{91}, 250401 (2003).

   \bibitem{ref:exp_Innsbruck}
      M.~Bartenstein, A.~Altmeyer, S.~Riedl \textit{et al.},
      Phys. Rev. Lett. \textbf{92}, 120401 (2004).

   \bibitem{unitarity}
      M.~Inguscio, W.~Ketterle, and C.~Salomon, eds.,
      \emph{Ultracold Fermi Gases}
      (IOS Press, Amsterdam, 2008),
      Proceedings of the International School of Physics ``Enrico Fermi,''
      Course CLXIV, Varenna, 20-30 June 2006.

   \bibitem{unitarity_new}
      S.~Riedl, E.R.~Sanchez Guajardo, C. Kohstall \textit{et al.},
      Phys. Rev. A \textbf{78}, 053609 (2008).

   \bibitem{carlson}
      J.~Carlson, S.-Y.~Chang, V.R.~Pandharipande \textit{et al.},
      Phys. Rev. Lett. \textbf{91}, 050401 (2003).

   \bibitem{ref:th}
      H.~Heiselberg,
      J. Phys. B \textbf{37}, S141 (2004).

   \bibitem{ref:leg80}
      A.J.~Leggett,
      in \emph{Modern Trends in the Theory of Condensed Matter},
      edited by A.~Pekalski and R.~Przystawa
      (Springer-Verlag, Berlin, 1980).

   \bibitem{ref:BCS1}
     J.R.~Engelbrecht, M.~Randeria, and C.A.R. S\'a de Melo,
     Phys. Rev. B \textbf{55}, 15153 (1997).

   \bibitem{ref:BCS2}
     M.M.~Parish, B.~Mihaila, E.M.~Timmermans \textit{et al.},
     Phys. Rev. B \textbf{71}, 064513 (2005).

   \bibitem{ref:CCE1}
      F.~Coester,
      Nucl. Phys. \textbf{7}, 421 (1958).

   \bibitem{ref:CCE2}
       F.~Coester and H.~K\"ummel,
       Nucl. Phys. \textbf{17}, 477 (1960).

   \bibitem{FC:lecture}
      F. Coester,
      \emph{The Ground State of Homogeneous Matter and the Foundation of the Nuclear Shell Model},
      in Lectures in Theoretical Physics, Quantum Fluids and Nuclear Matter,
      Vol XI B,
      edited by K.T. Mahanthapa and W.E. Brittin,
      (Gordon and Breach, New York, 1969).

   \bibitem{c-CCE}
      B.~Mihaila,
      Phys. Rev. C \textbf{68}, 54327 (2003).

   \bibitem{greiner}
      W. Greiner and A. Solov'yov,
      Chaos, Solitons and Fractals, 25, 835 (2005).
\end{thebibliography}
\end{document}